\documentstyle[prb,multicol,aps,psfig]{revtex}

\begin{document} 

\draft 
\tightenlines
\widetext

\title{On Metal--Insulator Transitions due to
Self--Doping} \author{Stefan Blawid, Hoang Anh Tuan\thanks{On leave
from National Institute of Physics, Hanoi, Vietnam}, Takashi
Yanagisawa\thanks{On leave from Electrotechnical Laboratory, Tsukuba,
Japan}~~and Peter Fulde} \address{Max--Planck--Institut f\"ur Physik
komplexer Systeme,\\ Bayreuther Str.\ 40, D-01187 Dresden, Germany}
\date{\today} \maketitle

\begin{abstract} We investigate the influence of an unoccupied band
on the transport properties of a strongly correlated electron
system.  For that purpose, additional orbitals are coupled to a
Hubbard model via hybridization. The filling is one electron per
site. Depending on the position of the additional band,
both, a metal--to--insulator and an insulator--to--metal transition
occur with increasing hybridization. The latter transition from a
Mott insulator into a metal via ``self--doping'' was recently
proposed to explain the low carrier concentration in $\rm Yb_4As_3$.
We suggest a restrictive parameter regime for this transition making
use of exact results in various limits. The predicted absence of the
self--doping transition for nested Fermi surfaces is confirmed by
means of an unrestricted Hartree--Fock approximation and an exact 
diagonalization study in one dimension. In the general case 
metal--insulator phase diagrams are obtained within the 
slave--boson mean--field and the alloy--analog approximation.
\end{abstract}


\begin{multicols}{2}

\narrowtext

\section{Introduction} \label{intro}

One of the most remarkable examples of the failure of the
independent--electron approximation is the existence of so--called
Mott--Hubbard insulators. The early transition oxides like $\rm
LaVO_3$, $\rm LaTiO_3$, $\rm V_2O_3$, $\rm Ti_2O_3$ or $\rm Cr_2O_3$
show an insulating behaviour, although they contain partially filled
$d$\/ bands. The reason is the strong Coulomb repulsion of the $d$\/
electrons, which hinders them from moving freely. A simple model to
describe the situation is the Hubbard Hamiltonian with one electron
per site \cite{Hubbard63a}:

\begin{equation} \label{hubbard} H\, = \, \sum_{i,j,\sigma} t_{ij} \,
f_{i\sigma}^{\,\dagger} f_{j\sigma}^{} + U \sum_i n_{i\uparrow}
n_{i\downarrow}.  \end{equation} Electrons can hop from one site $i$
to another $j$ via the hopping--matrix element $t_{ij}$. Whenever two
electrons (of opposite spins) occupy the same site, they repel each
other with an interaction energy $U$.

To obtain insulating behaviour (implying a vanishing dc conductivity
at zero temperature) an energy gap for charge excitations is
required. The existence of a conductivity gap can be inferred from
the spectral density of states $\rho (\omega)$. Obviously, for large
Coulomb repulsion and half--filling there is a gap of order $U$. The
transition from a metallic ($U=0$) to an insulating ($U$ large) state
is still subject of intensive investigations \cite{Georges95}. For
general lattices (without special symmetry properties) there is
reason to believe that the transition takes place at a finite value
$U= U_c \not= 0$ \cite{Georges95,Krishnamurthy90,Sorella92}.

The Hubbard model deals with one orbital per site and ignores any
additional bands which might be present. As was first pointed out by
Zaanen, Sawatzky and Allen \cite{Zaanen85} this is not sufficient for
a classification of the transition metal oxides. The following
Hamiltonian represents a natural extension of the Hubbard model by
including one additional spin--orbital with creation operators
$c_{i\sigma}^{\,\dagger}$ for each site $i$:

\begin{eqnarray} \label{extend} H & = & -\Delta \sum_{i\sigma}
n_{i\sigma}^{\,f} + \sum_{i,j,\sigma} t_{ij} \,
f_{i\sigma}^{\,\dagger} f_{j\sigma}^{} + U \sum_i n_{i\uparrow}^{\,f}
n_{i\downarrow}^{\,f} \nonumber \\
 & &{} + \sum_{i,j,\sigma} \tilde{t}_{ij} \, c_{i\sigma}^{\,\dagger}
c_{j\sigma}^{} + V \sum_{i\sigma} \left( f_{i\sigma}^{\,\dagger}
c_{i\sigma}^{} + {\rm h.c.} \right).  \end{eqnarray} In the following
we will refer sometimes to the additional orbital as a ligand orbital
implying that it may belong to a ligand atom. Note, that the on--site
energy of the ligand orbitals has been set equal to zero. Therefore
the ligand orbitals are higher in energy than the $f$\/ orbitals
($\Delta \geq 0$). The system is assumed to be quarter--filled, i.e.,
with one electron (or hole) per site.

Zaanen {\it et al}. \cite{Zaanen85} employed a single--impurity
approach to derive a phase diagram for the transition--metal
compounds. Thereby the transition--metal atoms are taken to be
independent. This implies a neglect of the hopping term $t_{ij}$ in
(\ref{extend}). Moreover, indirect hopping processes between the
$f$\/ orbitals via the $c$\/ orbitals are excluded. The calculated
band gaps in the density of states then approximately scale with the
hybridization $V$. Beside the Mott--Hubbard regime, where the gap is
proportional to $U$, a second region with a gap proportional to
$\Delta$ was identified. For this phase the term charge--transfer
insulator is used. As a consequence of the single--impurity approach
a metallic state was predicted when the transition--metal level lies
within the free electron band of the ligand host. In contrast, a
hybridization gap may be obtained when taking into account coherence
effects of the transition--metal atoms. Nimkar {\it et al}.
\cite{Nimkar93} and Sarma and Barman \cite{Sarma94} called the
insulating phase due to a hybridization gap a covalent insulator and
derived the corresponding region of the phase diagram. The special
feature of a covalent insulator is, that it undergoes an
insulator--to--metal transition with decreasing hybridization. The
gap does not scale with $V$. $\rm LnNiO_3$ (with Ln a rare earth
ion other than La) and $\rm NiS$ are believed to belong to this class
\cite{Nimkar93}. They can be driven into a metallic state without a
symmetry lowering structural transition by varying pressure or
temperature.

If we do not neglect the hopping term $t_{ij}$ in (\ref{extend}), we
can also imagine the opposite taking place, i.\/e., an
insulator--to--metal transition with increasing hybridization.
Electrons hopping into ligand levels create holes in the Hubbard
system, which may move and contribute to the conductivity. A
mechanism of this kind was recently proposed for the rare--earth
compound $\rm Yb_4As_3$ \cite{Fulde95}. Such an insulator--to--metal
transition due to ``self--doping'' would be an interesting new
possibility for forming metals with low carrier concentration.
Moreover, like the Mott--Hubbard insulator, it cannot be understood
within an independent electron picture.

To investigate in more detail the influence of an additional band on
the transport properties of a strongly correlated electron system we
suggest the following model:

\begin{eqnarray} \label{SMS} H & = & -\Delta \sum_{i,\sigma}
n_{i\sigma}^{\,f} + \sum_{i,j,\sigma} t_{ij} \,
f_{i\sigma}^{\,\dagger} f_{j\sigma}^{} + U \sum_i n_{i\uparrow}^{\,f}
n_{i\downarrow}^{\,f} \nonumber \\
& &{} + V \sum_{i,\sigma} \left(
f_{i\sigma}^{\,\dagger} c_{i\sigma}^{} + {\rm h.c.} \right).
\end{eqnarray} Again, the system is assumed to be quarter--filled,
i.\/e., it has one electron per site. The Coulomb repulsion is chosen
to be greater than the critical value $U_c$ of the Hubbard system.
Therefore we will call the model a self--doped Mott system (SMS) in
the following.  In comparison to the Hamiltonian~(\ref{extend}) we
have neglected the hopping $\tilde{t}_{ij}$ between the ligand
orbitals. For realistic situations this approximation is not
justified, neither for rare--earth nor for transition--metal
compounds. Note, that in systems like $\rm Yb_4As_3$ the widths 
of the correlated and uncorrelated bands are equal 
($t_{ij} = \tilde{t}_{ij}$). The Hamiltonian~(\ref{SMS}) should be 
regarded as a basic model containing the necessary terms for the 
self--doping mechanism. Moreover, whenever we have included the 
hopping term $\tilde{t}_{ij}$ we found no change in the general 
behaviour. We will come back to this point later.

In this paper we report first results of an investigation of the
system described by (\ref{SMS}). In Section~\ref{model} we illustrate
in a transparent way some model properties. We discuss exact
solutions in the limit of zero bandwidth and of infinite 
hybridization. Concerning the metal--insulator transition in the Hubbard 
model two different situations have to be distinguished. In the case 
of a nested Fermi surface the Hubbard model at half filling is an
insulator for any $U>0$ \cite{Hirsch85}. We investigate this case 
for the SMS by means of an unrestricted Hartree--Fock approximation 
\cite{Slater51} and present the results in Section~\ref{nefer}. 
Additionally the one--dimensional case is studied by exactly 
diagonalizing a finite system. Two early approaches have dominated 
the discussion about the Mott--Hubbard transition for general lattices 
without nesting: (i) the Brinkman--Rice approach \cite{Brinkman70},
which can be formally derived by using slave boson methods
\cite{Kotliar86}, and (ii) the alloy analogy, which is contained in
the first part of the well--known Hubbard III paper
\cite{Hubbard64}.  We apply these approximation schemes to our model
and present the results in Section \ref{genla}. Finally, 
Section~\ref{conclu} contains a short summary and outlook.

\section{The model} \label{model}

In the case of vanishing hybridization the SMS reduces to a Hubbard
model with a separated ligand level. In the first column of
Fig.~\ref{spectral} we sketch the corresponding density of states.
For $U > U_c$ three different regions exist (indicated by (2), (3)
and (4) in that figure):

\begin{figure}[htb]
\centerline{\psfig{figure=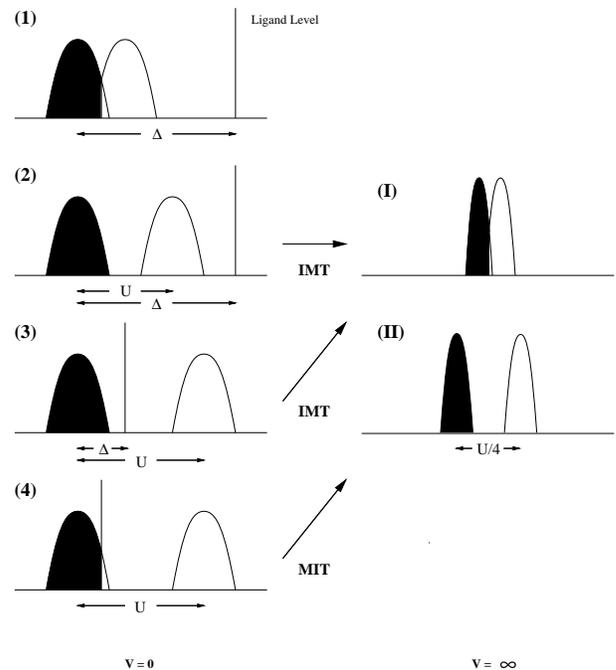,width=8cm,angle=0}}
\vspace{0.5cm}
\caption{\label{spectral}Schematic plot of the
density of state $\rho(\omega)$ for the SMS in the extreme limits of
vanishing and infinite hybridization. The occupied
(electron--removal) part is hatched. With IMT and MIT we indicate an
insulator--to--metal and a metal--to--insulator transition,
respectively. Different parameter regions are shown (see text).}
\end{figure}

\begin{itemize}

\item Mott--Hubbard insulator: $U < \Delta$. When adding one electron
to the system the lowest energy states contain an additional double
occupied $f$\/ orbital. The gap is of order $U$\/
\mbox{(Fig.~\ref{spectral}, (2))}.

\item Charge transfer insulator: $U > \Delta$. An additional electron
avoids the Coulomb repulsion felt in the $f$\/ subsystem and prefers
to occupy the ligand orbital. The energy scale of the gap is set by
$\Delta$ (Fig.~\ref{spectral}, (3)).

\item Due to the hopping term $t_{ij}$\/ the lower and upper Hubbard
bands possess a finite bandwidth $W(U)$. The ground state of the SMS
contains occupied ligand orbitals if the gain of kinetic energy due
to holes in the Hubbard subsystem overcomes the transfer energy:
$\Delta < W$. Adding or removing one electron changes the energy only
infinitesimally. The system is metallic (Fig.~\ref{spectral}, (4)).

\end{itemize}

We turn now to the case of infinite hybridization. For $V \gg
\Delta$\/ it is convenient to use a basis in which the hybridization
term is diagonal. Introducing:

\begin{equation} a_{i\sigma} =  \frac{1}{\sqrt{2}} \left( f_{i\sigma}
+ c_{i\sigma} \right) \; ,
 \; b_{i\sigma} =  \frac{1}{\sqrt{2}} \left( f_{i\sigma} -
 c_{i\sigma} \right), \end{equation} we obtain for the Hamiltonian
(\ref{SMS}):

\begin{equation} H = H_a + H_b + H_{ab} \end{equation} with:

\begin{eqnarray*} H_a & = & \left( -\frac{\Delta}{2} + V \right) \,
\sum_{i,\sigma}
	  a_{i\sigma}^{\,\dagger} a_{i\sigma}^{} + \frac{1}{2} \,
	  \sum_{i,j,\sigma} t_{ij} \, a_{i\sigma}^{\,\dagger}
	  a_{j\sigma}^{} \\
& &{}     + \frac{U}{4} \, \sum_{i} n_{i\uparrow}^a
	  n_{i\downarrow}^a \\
H_b & = & \left( -\frac{\Delta}{2} - V \right) \, \sum_{i,\sigma}
          b_{i\sigma}^{\,\dagger} b_{i\sigma}^{} + \frac{1}{2} \,
          \sum_{i,j,\sigma} t_{ij} \, b_{i\sigma}^{\,\dagger}
          b_{j\sigma}^{} \\ 
& &{}     + \frac{U}{4} \, \sum_{i} n_{i\uparrow}^b
          n_{i\downarrow}^b \\
\end{eqnarray*}

\begin{eqnarray*}
H_{ab} & = & - \frac{\Delta}{2} \, \sum_{i,\sigma} \left(
	     a_{i\sigma}^{\,\dagger} b_{i\sigma}^{} + {\rm h.c.}
	     \right) 
        + \, \frac{1}{2} \, \sum_{i,j,\sigma} t_{ij} \left(
	     a_{i\sigma}^{\,\dagger} b_{j\sigma} + {\rm h.c.} \right)\\
& &        - \, \frac{U}{4} \sum_{i} \{
             n_{i\uparrow}^{a} n_{i\downarrow}^a + n_{i\uparrow}^b
             n_{i\downarrow}^b  \\
& &     - \left.\left.\left.\left.\left.\left. \! \! \! \!
        \left(a_{i\uparrow}^{\,\dagger}+b_{i\uparrow}^{\,\dagger}
        \right) \right(a_{i\uparrow}^{}+b_{i\uparrow}^{} \right)
        \right(a_{i\downarrow}^{\,\dagger}+b_{i\downarrow}^{\,\dagger}
        \right) \right(a_{i\downarrow}+b_{i\downarrow} \right) \}.
\end{eqnarray*}

For $V = \infty$
the $a$\/ orbitals remain empty and we can restrict ourselves to the
part $H_b$. The SMS reduces to a Hubbard model with a hopping matrix
$t_{ij}/2$ and a Coulomb repulsion $U/4$. The result is easily
understood. When the two orbitals at a given site hybridize
strongly with each other, the eigenstates are given by the bonding
$b$\/ and \mbox{antibonding $a$\/} linear combinations. Only the
$b$\/ orbitals are populated to which the original $f$\/ orbitals
contribute by one half. The Coulomb repulsion felt by two electrons
occupying one $b$\/ orbital is reduced to a quarter and the hopping
between two $b$\/ orbitals is reduced to one half.  The trivial but
important observation is, that the Coulomb repulsion is stronger
reduced than the hopping. Since only the relation between these two
quantities is important, the SMS describes for $U/2 < U_c$ a metal
and for $U/2 > U_c$ an insulator ($V = \infty$). This is indicated in
the second column of Fig.~\ref{spectral} by (I) and (II),
respectively.

Having discussed the extreme limits, two transitions seem to be
possible when changing the hybridization $V$\/ from zero to
infinity:  For $\Delta < W(U)$, $U > 2 U_c$ a metal--to--insulator
and for $\Delta > W(U)$, $U_c < U < 2 U_c$ an insulator--to--metal
transition should take place with increasing $V$. Here $W(U)$\/ is
the effective band width of the lower Hubbard band. The two
possibilities correspond to the aforementioned covalent insulator and
a self--doped metallic state, respectively. It is worthwhile to note
that the second scenario is realized only if $U_c > 0$, i.\/e., for a
finite value of the critical Coulomb repulsion $U_c$\/ in the
corresponding Hubbard model.

In order to gain better insight into the behaviour of the system for
finite hybridization we have considered the zero bandwidth limit of
the SMS:

\begin{equation} H \, = \, -\Delta \sum_{\sigma} n_{\sigma}^{\,f} + U
\, n_{\uparrow}^{\,f} n_{\downarrow}^{\,f} + V \sum_{\sigma} \left(
f_{\sigma}^{\,\dagger} c_{\sigma}^{} + {\rm h.c.} \right).
\end{equation} The Hamiltonian is easily diagonalized, yielding the
eigenvalues $E_m^{(n)}$\/ and eigenstates
$\left|m^{(n)}\,\right\rangle$\/ for $n$\/ electrons ($n = 1,2,3$\/).
The $f$\/ density of states is given by:

\begin{eqnarray} \label{spec} \rho^{\,f}(\omega) & = &
\sum_{m,i,\sigma}  \{ \,
		   \left| \left\langle\,  m^{(N_e+1)} \left|
		   f_{i\sigma}^{\,\dagger} \right| 0^{(N_e)}
		   \right\rangle \right|^2 \nonumber \\
& &{} \;\;\;\;\;\;\;\;\,\;\;\;\;\;
      \times \delta(\omega+E_0^{(N_e)}-E_m^{(N_e+1)}) \, 
      \nonumber \\
& &   \;\;\;\;\;\; + \left| \left\langle\, m^{(N_e-1)}
      \left|f_{i\sigma}^{}\right| 0^{(N_e)} \right\rangle \right|^2 
      \nonumber \\
& &   \;\;\;\;\;\;\;\;\,\;\;\;\;\; 
      \times \delta(\omega+E_m^{(N_e-1)}-E_0^{(N_e)}) \}.
\end{eqnarray} Here $N_e = 1$\/ and the sum is only over one site.
Furthermore, $\mid\!0^{(N_e)}\rangle$ denotes the ground state. A
similar expression holds for the density of states of the $c$\/
electrons.  The total one is given by $\rho(\omega) = \rho^f(\omega)
+ \rho^c(\omega)$.

\begin{figure}[htb]
\psfig{figure=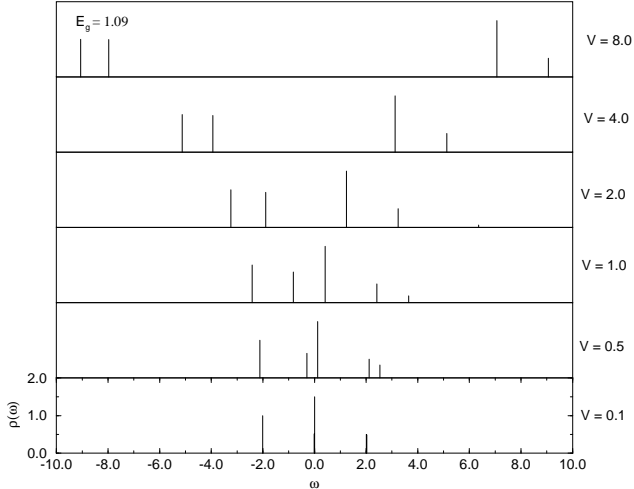,width=8cm,angle=-90}
\vspace{0.5cm}
\caption{\label{evolution}Schematic representation of
the excitations from the ground state of the SMS in the zero
bandwidth limit on adding and removing an electron. Indicated are the
weights in the spectral density.  Various values of the hybridization
are used ($\Delta = 2$, $U = 4$).}
\end{figure}

In Fig.~\ref{evolution} we show the results of the diagonalization.
Plotted are the positions of the poles and their weights in the 
density of states. For small hybridization we can clearly identify 
the lower and upper Hubbard peaks, as well as the peaks caused by 
adding one electron to the ligand orbital situated at 
$\omega \approx 0$ 
($f^{\,1}c^{\,0} \rightarrow f^{\,1}c^{\,1}$) and $\omega \approx
\Delta$ ($f^{\,1}c^{\,0} \rightarrow f^{\,0}c^{\,2}$). Due to the
different effect of the hybridization on a singlet and a triplet
state the \mbox{$f^{\,1}c^{\,1}$--peak} splits up. As a result weight
is pushed towards the lower Hubbard peak. We stress that there is no
direct weight transfer between the lower and upper Hubbard peak.
Instead, the weight of the lower Hubbard peak is always equal to
unity. Thus a simple self--doping picture which predicts a metallic
state for an infinitesimal small hybridization $V$ does not hold
true in that case. Note, that the gap 
$E_{\rm gap} = E_0^{(2)} + E_0^{(0)} - 2 E_0^{(1)}$ between the 
occupied and unoccupied part of the spectra
continously decreases from $E_{\rm gap} = \Delta$ to 
$E_{\rm gap} \approx U/4$. When the full Hamiltonian is used the 
peaks will mark the position of centers of bands. An overlap
of the bands accompanied by an insulator--to--metal transition is 
expected, when the hybridization $V$\/ exceeds a critical value 
$V_c$. 

\section{Nested Fermi surface} \label{nefer}

We turn now to a treatment of the full Hamiltonian (\ref{SMS}). In 
this section we consider lattice symmetries and hopping matrices 
which lead to a nested Fermi surface. The critical Coulomb repulsion 
in the corresponding Hubbard model vanishes. In one spatial dimension
this is the general behaviour. The dimension $d=1$ allows for an
exact diagonalization study. Higher dimensions are treated within the
unrestricted Hartree--Fock approximation, which getting exact in 
the limit $U \rightarrow 0$ and in the atomic limit $U = \infty$.
Note, that in the case of $U_c = 0$ small Coulomb repulsions are the
most interesting. After the discussion of Section~\ref{model} the 
behaviour for values 
\mbox{ $U$ \hspace*{-1.5ex} 
\raisebox{-0.3ex}{ $\stackrel{\scriptstyle>}{\scriptstyle\sim}$ } 
\hspace*{-1.5ex} $U_c$ }
should be representative also for all $U > U_c$.

\subsection{\label{sechf}Unrestrictive Hartree--Fock approximation}

We assume in the following a bipartite lattice. Additional lattice 
symmetry properties enter the calculation only via the band
dispersion $\epsilon_{\vec{k}}$\/, which is the Fourier transform of
the hopping matrix

\begin{equation} t_{ij} = \frac{1}{N} \, \sum_{\vec{k}}
\epsilon_{\vec{k}} \, e^{i \, \vec{k} \, (\vec{r}_i - \vec{r}_j)}.
\end{equation} Here $N$\/ is the number of sites in the lattice. It
is convenient to introduce the density of states of free particles:

\begin{equation} \rho_0(\omega) = \frac{1}{N} \, \sum_{\vec{k}}
\delta(\omega-\epsilon_{\vec{k}}).  \end{equation} In the
calculations presented in this paper it suffices to specify
$\rho_o(\omega)$\/ instead of the lattice structure and the hopping
matrix elements. In order to proceed analytically as far as possible,
we will use in the following a rectangular density of states:

\begin{equation} \rho_o(\omega) = \frac{1}{2 W} \, \Theta(W-\omega)
\, \Theta(W+\omega), \end{equation} where $\Theta(\omega)$ is the
step function. This choice maintains the antiferromagnetic instability
for arbitrarily small values of $U$ and represents the situation of
a nested Fermi surface.

In the mean--field approach the interaction term of the Hamiltonian
(\ref{SMS}) is linearized, i.\/e.,

\begin{equation} n_{i\uparrow}^f n_{i\downarrow}^f \, \longrightarrow
\, n_{i\uparrow}^f \langle\,n_{i\downarrow}^f\,\rangle +
\langle\,n_{i\uparrow}^f\,\rangle n_{i\downarrow}^f -
\langle\,n_{i\uparrow}^f\,\rangle
\langle\,n_{i\downarrow}^f\,\rangle.  \end{equation} The electrons
interact via a molecular field. Depending on the anticipated symmetry
of the ground state, different ansatzs for the expectation values
$\langle\,n_{i\sigma}^{\,f}\,\rangle$\/ are made.  In the case of
antiferromagnetic order two sublattices $A$\/ and $B$ are introduced.
The following ground states are considered\\[1ex]
\begin{tabular}{clcl} $\bullet$ & paramagnetic & : &
$\langle\,n_{i\uparrow}^f\,\rangle = \frac{1}{2} \, (1-\delta) =
\langle\,n_{i\downarrow}^f\,\rangle$, \\[1.5ex] 
$\bullet$ &
ferromagnetic & : & $\langle\,n_{i\uparrow}^f\,\rangle = \frac{1}{2}
\, (1-\delta+m)$, \\[1.5ex] & & & 
$\langle\,n_{i\downarrow}^f\,\rangle = \frac{1}{2}
\, (1-\delta-m)$, \\[1.5ex] 
$\bullet$ & antiferromagnetic & : &
$\langle\,n_{i\uparrow}^{f\,A}\,\rangle = \frac{1}{2} \, (1-\delta+m)
= \langle\,n_{i\downarrow}^{f\,B}\,\rangle$, \\[1.5ex] & & &
$\langle\,n_{i\downarrow}^{f\,A}\,\rangle = \frac{1}{2} \,
(1-\delta-m) = \langle\,n_{i\uparrow}^{f\,B}\,\rangle$ .
\end{tabular}\\[1ex] These are the main states in the phase diagram of the
doped Hubbard model calculated by Penn \cite{Penn65}. We indicate now
the construction of the phase diagram for the SMS. Following Langer
{\it et al}. \cite{Langer69} we solve the equation of motion for the
one--particle Green's function $\langle\langle\,f_{\vec{k}\sigma}\, ;
\, f_{\vec{k}\sigma}^{\,\dagger}\,\rangle\rangle \! {\atop\omega}$
and $\langle\langle\,c_{\vec{k}\sigma}\, ; \,
c_{\vec{k}\sigma}^{\,\dagger}\,\rangle\rangle \! {\atop\omega}$. From
the imaginary part the spectral densities
$\rho_{\vec{k}\sigma}^f(\omega)$\/ and
$\rho_{\vec{k}\sigma}^c(\omega)$\/ are obtained. For zero temperature
the \mbox{magnetization $m$} and the self--doping $\delta$\/ are
given by the solution of:

\begin{eqnarray} \label{hf} \frac{1}{N}\,\sum_{\vec{k},\sigma} \,
\int\limits_{-\infty}^{\mu} \left( \rho_{\vec{k}\sigma}^f(\omega) +
\rho_{\vec{k}\sigma}^c(\omega) \right) \, d\omega
 & = & 1 \nonumber \\ \frac{1}{N}\,\sum_{\vec{k}} \,
\int\limits_{-\infty}^{\mu} \rho_{\vec{k}\uparrow}^f(\omega)  \,
d\omega
 & = & \frac{1}{2} \left(1-\delta+m \right) \\
\frac{1}{N}\,\sum_{\vec{k}} \, \int\limits_{-\infty}^{\mu}
\rho_{\vec{k}\downarrow}^f(\omega)  \, d\omega
 & = & \frac{1}{2} \left(1-\delta-m \right) \, , \nonumber
\end{eqnarray} where we have introduced the chemical potential $\mu$.
The equations (\ref{hf}) are solved numerically by iteration using
for $\rho_0(\omega)$ a rectangular form. Knowing the parameters $m$,
$\delta$ and $\mu$\/ for the ground state of a given symmetry the
energies of the various states are calculated. The state with the
lowest energy is the true ground state and determines the spectrum.

First we discuss the result for $V=0$. In the antiferromagnetic 
phase the model corresponds to an insulator. The corresponding spectral 
density is the one indicated in Fig.~\ref{spectral}, (2) and (3). 
For $\Delta > W$\/ the ground state is antiferromagnetic for all $U$. 
This phase can also be present if $\Delta < W$. In particular we obtain 
an insulating antiferromagnetic solution for small $U < \Delta$\/ up
to $\Delta = 0$. This reflects the fact that the ground state does not 
contain occupied $c$\/ orbitals. The $f$\/ subsystem is undoped. 

We now turn to the results for finite hybridization. Our calculations 
show that the insulating phases survive, when the hybridization $V$
is changed from zero to a finite value. In the limit 
$V \rightarrow \infty$\/ the ground state is an antiferromagnetic 
insulator for all values of $\Delta$ and $U$. No insulator--to--metal
transition takes place. In Section~\ref{model} we have argued that 
this behaviour requires a vanishing critical Coulomb repulsion for 
the Hubbard model. Therefore the obtained result confirms our 
prediction.

As mentioned before, in the limit $V \rightarrow \infty$\/ the system 
is always insulating. This finding implies a metal--to--insulator 
transition due to the opening of a hybridization gap in a certain 
parameter regime. We have identified a regime with $V_c = 0$ and one 
with a finite critical value of the hybridization for the transition
to an insulator. Details will be presented elsewhere. Here we 
characterize the two regimes only roughly by 
$\Delta \leq U \leq 2 W$\/ and $2 W \leq U \leq 4 W$, respectively
($\Delta \ll W$).

\subsection{\label{secla}Exact diagonalization in one dimension}

In one spatial dimension the critical Coulomb repulsion $U_c$\/ of
the Hubbard model with nearest neighbour hopping vanishes. The
model should be insulating for all values of $U$\/ and $V$\/
( $\Delta > W$). Extending the Hamiltonian (\ref{SMS}) to 
(\ref{extend}), i.\/e., including a hopping term between different 
$c$\/ orbitals, does not change the result qualitatively. In the 
limit $V \rightarrow \infty$ the Hamiltonian (\ref{extend}) reduces 
to a single band Hubbard model with a Coulomb repulsion $U/4$ and 
hopping matrix elements $(t_{ij} + \tilde{t}_{ij})/2$. A self-doping 
transition can only be expected within the parameter regime 
$\Delta > W(U) + \tilde{W}$ and  
$U_c < U < 2\,(1 + \tilde{W}/W)\,U_c$. With $W$\/ and $\tilde{W}$\/ we 
denote half of the bandwidth of the $f$\/ and the $c$\/ bands,
respectively. Here the conditions on $U$\/ are less restrictive as
before. 

To verify the prediction we have performed an exact 
diagonalization of the Hamiltonian (\ref{extend}) on a four site chain. 
Treating the extended model in this approach is no additional difficulty.
Before presenting the results we first indicate the calculations. The 
one particle excitation spectrum is defined by Eq.\/~(\ref{spec}). 
The first and second term give the electron addition and electron 
removal spectrum, respectively. The spectral density $\rho(\omega )$ 
for small clusters can be obtained by the Lanczos algorithm in form of 
a continued-fraction \cite{Dagotto94}. In order to obtain the electron 
removal spectrum, we start from the vector 
$\psi_0 = f_{i \sigma}\,\psi_0^{(Ne)}$\/ and then tridiagonalize the 
Hamiltonian by the Lanczos method. The positions
of the poles and their weights follow from diagonalizing the
tridiagonal matrix. The electron addition spectrum is obtained
similarly.

\begin{figure}[htb]
\psfig{figure=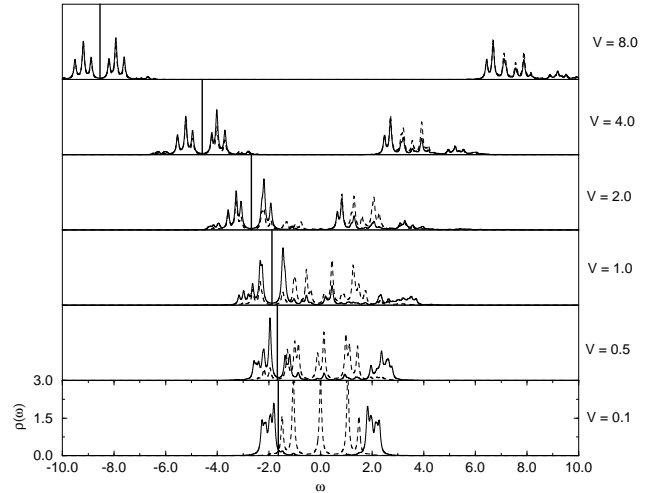,width=8cm,angle=-90}
\vspace{0.5cm}
\caption{\label{lanczos}Evolution of the spectral
density of the extended model (\protect{\ref{extend}}) in one
dimension. Results are shown for a four site chain. Periodic and
antiperiodic boundary conditions are used. The parameters are given
by: $U=4$, $\Delta=2$, $W=0.25$, $\tilde{W}=1.5$. The solid and
dashed lines indicate the $f$\/ and $c$\/ spectral density,
respectively. The vertical line marks the position of the chemical
potential.  The peaks have been broadened by $\epsilon=0.05$.}
\end{figure}

Results are shown in Fig.~\ref{lanczos}. We follow the usual
representation and plot the $\delta$--function of
equation~(\ref{spec}) with a (arbitrary) width $\epsilon = 0.05$
\mbox{ $(\delta( \omega - \omega_{\rm pol}) \rightarrow
-\frac{1}{\pi} \, {\rm Im} \left( \frac{1}{\omega - \omega_{\rm pol}
+ i \epsilon} \right) )$}.  A pseudo--gap seems to exist between the
electron removal and the electron addition spectra. This is caused by
the broadening. In all our calculations there is clearly a gap
between the occupied and unoccupied part of the spectrum. The energy
regime without weight is much larger than the typical distance
between the poles. To summarize, the Lanczos calculations predict an
insulator for the extended model $(\Delta > W + \tilde{W})$ in one
spatial dimension. Thus in the extended model the self-doping
transition is restricted to a limited parameter regime. This
behaviour is similar to the one of the SMS.

\section{General lattices} \label{genla}

In this section we treat the Hamiltonian~(\ref{SMS}) within (i) the 
slave--boson mean--field and (ii) the alloy--analog approximation.
We restrict ourselves to a paramagnetic ground state. Nevertheless
correlation effects are included and both approximations predict a 
finite critical Coulomb repulsion for the Mott--Hubbard transition. 
The restriction to a paramagnetic ground state does not allow the
resolution of details of the underlying lattice symmetry. The 
results are therefore interpreted as the behaviour of the SMS on a 
general lattice. We have excluded here the case of a nested Fermi 
surface. Again it suffices to specify $\rho_o(\omega)$. In the 
slave--boson approach we will use a rectangular form as before and 
in the alloy analogy we will choose an elliptical one:

\begin{equation} \rho_o(\omega) = \frac{2}{W^2 \pi} \,
\sqrt{W^2-\omega^2}.  \end{equation}

\subsection{\label{secsb}Slave--boson mean--field approximation}

In the slave--boson approach four bosonic fields are introduced with
creation operators $d_i^{\,\dagger}$, $s_{i\sigma}^{\,\dagger}$\/ and
$e_i^{\,\dagger}$. They correspond to the four possible states of an
$f$\/ orbital, i.\/e., double occupation, single occupation with spin
$\sigma = \pm 1$\/ and empty state. The following constrains ensure
that only the physically relevant part of the enlarged Hilbert space
is considered

\begin{eqnarray} \label{constrain} e_i^{\,\dagger}e_i^{} +
s_{i\uparrow}^{\,\dagger}s_{i\uparrow}^{} +
s_{i\downarrow}^{\,\dagger}s_{i\downarrow}^{} + d_i^{\,\dagger}d_i^{}
& = & 1 \nonumber \\ s_{i\sigma}^{\,\dagger}s_{i\sigma}^{} +
d_i^{\,\dagger}d_i^{} & = & f_{i\sigma}^{\,\dagger}f_{i\sigma}^{}.
\end{eqnarray} Having introduced the bosonic fields the Hamiltonian
(\ref{SMS}) projected onto the physical subspace can be expressed in
the form

\begin{eqnarray} \label{sbham} H & = & -\Delta \sum_{i,\sigma}
f_{i\sigma}^{\,\dagger}f_{i\sigma}^{} + \sum_{i,j,\sigma} t_{ij} \,
z_{i\sigma}^{\,\dagger} f_{i\sigma}^{\,\dagger} f_{j\sigma}^{}
z_{j\sigma}^{} + U \sum_i d_i^{\,\dagger}d_i^{} \nonumber \\
& &{} + V \sum_{i,\sigma}
\left( z_{i\sigma}^{\,\dagger} f_{i\sigma}^{\,\dagger} c_{i\sigma} +
{\rm h.c.} \right), \end{eqnarray} where $z_{i\sigma}^{} =
e_i^{\,\dagger}s_{i\sigma}^{} + s_{i-\sigma}^{\,\dagger}d_i^{}$. As
long as the constrains (\ref{constrain}) are satisfied, the following
replacement is possible:

\begin{eqnarray} \label{z} z_{i\sigma}^{} \longrightarrow
\tilde{z}_{i\sigma}^{} & = & 
\frac{1}{\sqrt{1-d_i^{\,\dagger}d_i^{}-s_{i\sigma}^{\,\dagger}s_{i\sigma}^{}}}
\, z_{i\sigma}^{} \nonumber \\
& &{} \;\;\;\;\;\;\;\;\;\;\;\;\;\; \times
\frac{1}{\sqrt{1-e_i^{\,\dagger}e_i^{}-
s_{i-\sigma}^{\,\dagger}s_{i-\sigma}^{}}}.  \end{eqnarray} Equations
(\ref{constrain}), (\ref{sbham}) and (\ref{z}) provide an exact
representation of the projected original Hamiltonian. In the
mean--field approximation the different representations
$z_{i\sigma}^{}$\/ and $\tilde{z}_{i\sigma}^{}$\/ are not identical
any more. The form chosen in (\ref{z}) gives the correct result for
the Hubbard model in the limit $U = 0$.

In order to perform the mean--field approximation in the paramagnetic
phase, we have replaced the operators by real, spin-- and
site--independent expectation values $e$\/, $s$\/ and $d$. Due to the
constrains only one of these values is independent, say $d^{\,2}$
($=b$). The Hamiltonian can then be written in the form:

\begin{eqnarray} \label{mfsbham} H & = & -\Delta \sum_{i,\sigma}
f_{i\sigma}^{\,\dagger}f_{i\sigma}^{} + \tilde{z}^2 \sum_{i,j,\sigma}
t_{ij} \, f_{i\sigma}^{\,\dagger} f_{j\sigma}^{} + U N b \nonumber \\
& &{} + V \tilde{z} \sum_{i,\sigma} \left( f_{i\sigma}^{\,\dagger} c_{i\sigma}
+ {\rm h.c.} \right), \end{eqnarray} where

\[ \tilde{z}^{\,2} = \frac{(n_f-b)(\sqrt{b}+\sqrt{1+b-2
n_f})^{\,2}}{(1-n_f)\,n_f}, \] ($ n_f =
\langle\,n_{\uparrow}^f\,\rangle =
\langle\,n_{\downarrow}^f\,\rangle{} $).  The mean--field approach
yields an insulating state for vanishing $\tilde{z} = 0$. In that
case the renormalized hopping and hybridization vanish. All electrons
occupy $f$\/ orbitals and are localized. To calculate the unknown
parameters $n_f$ and $b$ we proceed in a similar way as described in
Section (\ref{sechf}). The equations of motion are solved for the
one--particle Green's functions $\langle\langle\,f_{\vec{k}\sigma}\,
; \, f_{\vec{k}\sigma}^{\,\dagger}\,\rangle\rangle \! {\atop\omega}$,
$\langle\langle\,c_{\vec{k}\sigma}\, ; \,
c_{\vec{k}\sigma}^{\,\dagger}\,\rangle\rangle \! {\atop\omega}$\/ and
$\langle\langle\,c_{\vec{k}\sigma}\, ; \,
f_{\vec{k}\sigma}^{\,\dagger}\,\rangle\rangle \!
{\atop\omega}$\/ and the corresponding spectral densities are
calculated. The following conditions must be obeyed:

\begin{eqnarray} \label{sb} \frac{1}{N}\,\sum_{\vec{k},\sigma} \,
\int\limits_{-\infty}^{\mu} \left( \rho_{\vec{k}\sigma}^f(\omega) +
\rho_{\vec{k}\sigma}^c(\omega) \right) \, d\omega
 & = & 1 \nonumber \\ \frac{1}{N}\,\sum_{\vec{k},\sigma} \,
\int\limits_{-\infty}^{\mu} \rho_{\vec{k}\sigma}^f(\omega)  \,
d\omega
 & = & 2\,n_f \\ \langle\, \frac{\partial H}{\partial b} \,\rangle =
\frac{\partial (\tilde{z}^{\,2})}{\partial b} \,
\sum_{\vec{k},\sigma} \, \int\limits_{-\infty}^{\mu}
\epsilon_{\vec{k}}^{} \, \rho_{\vec{k}\sigma}^f(\omega) \, d\omega \,
+ \, U N &  & \nonumber \\
+ \, 2 V \, \frac{\partial \tilde{z}}{\partial
b} \, \sum_{\vec{k},\sigma} \, \int\limits_{-\infty}^{\mu}
\rho_{\vec{k}\sigma}^{fc}(\omega) \, d\omega
 & = & 0 \, .\nonumber \end{eqnarray} We have solved equations (\ref{sb})
by using for $\rho_0(\omega)$ a rectangular form. An analytic
expression for the surface $\tilde{z} = 0$ in the ($U$,$\Delta$,$V$)
space is obtained, starting from the metallic regime.

\begin{figure}[htb]
\centerline{\psfig{figure=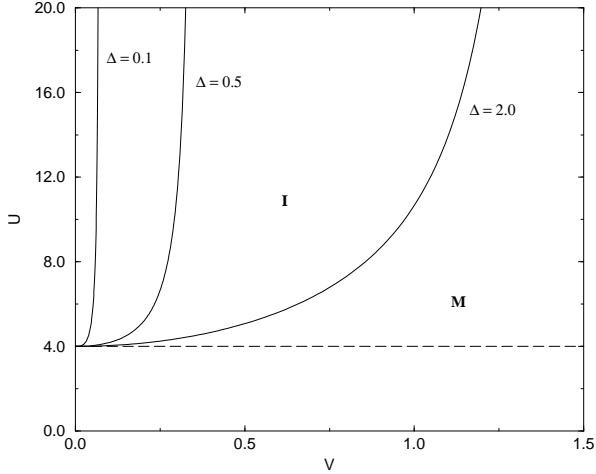,width=9cm,angle=-90}}
\vspace{0.5cm}
\caption{\label{sbfig}Phase diagram for the SMS in the slave--boson
mean--field approximation. The solid lines separate insulating and
metallic regimes. The dashed line indicates the Brinkman--Rice value
for the critical Coulomb repulsion.
Below the line the system is
metallic for all values of $\Delta$\/ and $V$\/ ($W = 1$). In the limit 
$U \rightarrow \infty$ a metallic phase exists for 
$V > V_c = \Delta / \protect{\sqrt{2}}$
.}
\end{figure}

For $V = 0$\/ we recover the well known result of Brinkman and Rice
\cite{Brinkman70}. The critical Coulomb repulsion for the
Mott--Hubbard transition is found to be $U_{\rm BR} = 4 W$. The value
is independent of $\Delta$ even for $\Delta < W$, because the
renormalized bandwidth reduces to zero in the insulating phase. For
$U < U_{\rm BR}$\/ a metallic state exist for all values of $V$. For
$U > U_{\rm BR}$ an insulator--to--metal transition is obtained at a
critical value of the hybridization given by:

\begin{equation} V_c \, = \, \Delta \left( \frac{\sqrt{4 \Delta^2+U
(U-4 W)} - 2 \Delta}{2 U} \right)^{\frac{1}{2}}.  \end{equation} The
resulting phase diagram is shown in Fig.~\ref{sbfig}. To summarize,
the slave--boson approach supports the existence of an
insulator--to--metal transition with increasing hybridization. In
contrast to the arguments given in Section \ref{model} the transition
is also found for \mbox{$U > 2 U_{\rm BR}$}. The transition takes
place at a finite critical value of the hybridization which in the
limit $U \rightarrow \infty$ is given by $V_c = \Delta/\sqrt{2}$.

We have also applied the slave--boson mean--field approximation to the 
Hamiltonian (\ref{extend}). For $\Delta > \tilde{W}$ the result is 
similar to the one obtained for the SMS. As before we denote with 
$\tilde{W}$\/ half of the bandwidth of the $c$\/ band. The expression 
for the critical hybridization becomes here:

\begin{equation} V_c = \left\{ \begin{array}{c@{\quad,\quad}ll} 0 &
\mbox{if $U < 4 W$} \\ \left( \frac{\alpha}{U} \, \frac{\sqrt{4
\alpha^2\beta^2 + U (U-4 W)} - 2 \alpha \beta}{2}
\right)^{\frac{1}{2}} & \mbox{otherwise} \end{array} \right.
\end{equation} where $\alpha = \Delta\,( \Delta - \tilde{W})$ ,
$\beta = \frac{1}{\tilde W}\, \ln{\frac{\Delta}{\Delta - \tilde
W}}$.  As before in the SMS, a self-doping transition takes place for
all values $U > U_{\rm BR}$.

\subsection{\label{seccpa}Alloy--analog approximation}

As first pointed out by Velick\'{y} {\it et al}. \cite{Velicky68} the
``scattering correction'', described in \cite{Hubbard64}, can be
viewed as an alloy problem. Applying the approximation to the SMS we
follow the work of Ref.\/ \cite{Leder79} for the periodic Anderson
model.  Alternatively, a Green's function decoupling scheme in the
spirit of the ``scattering correction'' can be applied. Both
approaches are equivalent. The alloy--analog approach starts from an
intuitive physical picture. An electron with spin $\sigma$ can only
hop onto the $f$\/ orbital situated at site $i$, if the orbital is
either empty or occupied by an electron with spin $-\sigma$. The
electron with spin $\sigma$ can therefore be thought of as moving in
a static random potential with eigenvalues and probabilities given by

\begin{equation} \label{random} E_{i\sigma} = \left\{
\begin{array}{c@{\quad,\quad}ll} -\Delta & \mbox{with probability} &
1-\langle\,n_{i-\sigma}^f\,\rangle \\ -\Delta+U & \mbox{with
probability} & \langle\,n_{i-\sigma}^f\,\rangle \end{array} \right.
\, .  \end{equation} The many--body Hamiltonian (\ref{SMS}) is
replaced by a one particle one with disorder and is of the form:

\begin{eqnarray} \label{disham} H & = & \sum_{i,\sigma} E_{i\sigma}
\, f_{i\sigma}^{\,\dagger} f_{i\sigma}^{} + \sum_{i,j,\sigma} t_{ij}
\, f_{i\sigma}^{\,\dagger} f_{j\sigma}^{} \nonumber \\
& &{} + V \sum_{i,\sigma} \left(
f_{i\sigma}^{\,\dagger} c_{i\sigma}^{} + {\rm h.c.} \right) \, .
\end{eqnarray} In the following we assume site-- and
spin--independent expectation values in (\ref{random}) ($ n_f =
\langle\,n_{i\uparrow}^f\,\rangle =
\langle\,n_{i\downarrow}^f\,\rangle $).  The Green's function $G$\/
corresponding to the Hamiltonian~(\ref{disham}) has to be averaged
over all possible configurations of the random potential which can be
considered being due to impurities. The averaging cannot be performed
exactly. To solve the alloy problem the coherent potential
approximation (CPA) is used. The averaged Green's function
$\bar{G}$\/ follows from an effective Hamiltonian containing a
dynamical mean--field $\Sigma(\omega)$:

\begin{eqnarray} H_{\rm eff} & = & \Sigma(\omega) \,
\sum_{i,\sigma} \, f_{i\sigma}^{\,\dagger} f_{i\sigma}^{} +
\sum_{i,j,\sigma} t_{ij} \, f_{i\sigma}^{\,\dagger} f_{j\sigma}^{}
\nonumber \\
& &{} + V \sum_{i,\sigma} \left( f_{i\sigma}^{\,\dagger} c_{i\sigma}^{} +
{\rm h.c.} \right).  \end{eqnarray} $\Sigma(\omega)$ will be
determined in (\ref{sig}). For an elliptic density of states
$\rho_0(\omega)$\/ the averaged $f$\/ Green's function is therefore
of the form:

\begin{eqnarray} \label{avgree} \bar{G}^{ff}(\omega) & = &
\frac{1}{N} \, \sum_{\vec{k}}
\overline{\langle\langle\,f_{\vec{k}\sigma}\, ;
 \, f_{\vec{k}\sigma}^{\,\dagger}\,\rangle\rangle \! {\atop\omega}}
\nonumber \\ 
& = & \frac{2}{W^{\,2}} \, ( \left(\omega -
\frac{V^{\,2}}{\omega} - \Sigma(\omega) \right) \nonumber \\
& &{} \;\;\;\;\;\;\; - \sqrt{ \left(\omega
- \frac{V^{\,2}}{\omega} - \Sigma(\omega) \right)^2 - W^{\,2}} \; ).  
\end{eqnarray} The other Green's functions,
$\bar{G}^{fc}$\/ and $\bar{G}^{cc}$, can be expressed in terms of
$\bar{G}^{ff}$. A scattering matrix $T$\/ is introduced for each
configuration via:

\begin{equation} G \, = \, \bar{G} + \bar{G}\,T\,\bar{G}.
\end{equation} The dynamical mean--field is determined by demanding
that the scattering matrix vanishes on average: $\bar{T} = 0$. This
yields an expression for $\Sigma(\omega)$ of the form:

\begin{equation} \label{sig} \Sigma(\omega) \, = \, \bar{E} - \left(
-\Delta-\Sigma(\omega)  \right) \left( -\Delta + U - \Sigma(\omega)
\right) \, \bar{G}^{ff}(\omega), \end{equation} where $\bar{E} =
-\Delta + U \, n_f$. Eliminating $\Sigma(\omega)$\/ from
(\ref{avgree}) and (\ref{sig}) gives a cubic equation for
$\bar{G}^{ff}(\omega)$. By guessing a starting value for $n_f$\/ one
solves this equation and calculates the $f$\/ and the related $c$\/
density of states $\bar{\rho}_{}^{\,f}(\omega)$ and
$\bar{\rho}_{}^{\,f}(\omega)$, respectively.  An improved value of
$n_f$\/ is obtained from the solution of:

\begin{equation} \begin{array}{rcl} {\displaystyle
\int\limits_{-\infty}^{\mu} \bar{\rho}_{}^{\,f}(\omega) \, d\omega }
& = & {\displaystyle n_f } \\ {\displaystyle
\int\limits_{-\infty}^{\mu} \left( \bar{\rho}_{}^{\,f}(\omega) +
\bar{\rho}_{}^{\,c}(\omega) \, \right) d\omega } & = & {\displaystyle
\frac{1}{2} }.  \end{array} \end{equation} The equations are iterated
until self--consistency is obtained.

\begin{figure}[htb]
\psfig{figure=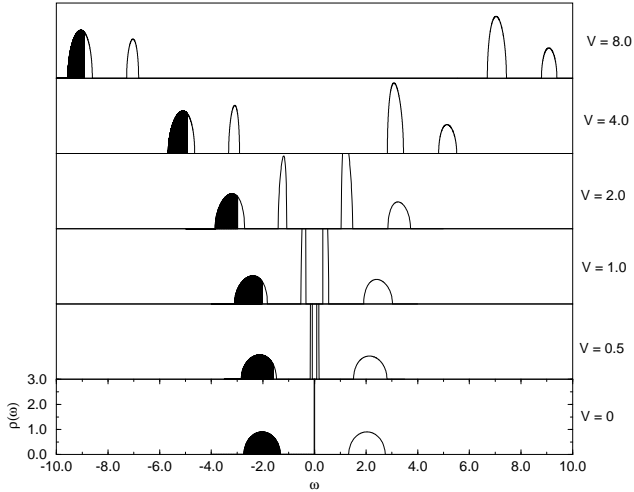,width=8cm,angle=-90}
\vspace{0.5cm}
\caption{\label{cpa}Evolution of the spectral density
with increasing hybridization for the SMS in the alloy--analog
approximation. The filled area indicates the occupied part of the
spectra. The used parameters are given by: $U=4$, $\Delta=2$, $W=1$.}
\end{figure}

It is well known that the alloy analogy gives a critical Coulomb
repulsion of \mbox{$U_{\rm AA} = W$\/} for the metal--insulator
transition in the Hubbard model. A typical result for \mbox{$U >
U_{\rm AA}$\/} is shown in Fig.~\ref{cpa}. The total weight of the
lower Hubbard band is given by \mbox{$2 \, (1-n_f)$}, independent of
the on--site energy of the ligand orbitals. For any finite value of
the hybridization, $n_f$\/ is less than $1/2$ and the system is
metallic. This corresponds to the simple self--doping argument
mentioned in Section~\ref{intro} and~\ref{model}. The obtained weight
transfer from the upper to the lower Hubbard band is not surprising,
since the expectation value $n_f$\/ determines the probability for
the two on--site energies $-\Delta$\/ and $-\Delta+U$ of the $f$\/
orbitals in the alloy problem (Eq.\/ (\ref{random})). The comparison
of Fig.~\ref{evolution} and \ref{cpa} reveals, that the overall
evolution of the spectrum is well described within the alloy--analog
approximation. The exact peaks obtained in the limit of zero
bandwidth are situated in the middle of the alloy bands. The only
exception is the band which is the singlet $f^{\,1}c^{\,1}$--peak in
the zero bandwidth limit. It is shifted toward higher energies. In
the limit $W \rightarrow 0$ and $V \rightarrow \infty$ the gap
between the lowest two peaks is given in the alloy approach by
$U/2$\/ and not by $U/4$\/ as in the exact result.

\section{Conclusion} \label{conclu}

We have investigated the effect of a hybridization $V$\/ between an
orbital with strong Coulomb repulsion and a ligand orbital on the
transport properties of a Hubbard system. The Coulomb repulsion $U$\/
was taken to be larger than the critical interaction $U_c$ for the
Mott-Hubbard transition of a system without ligand orbitals,
i.\/e., $U> U_c$. The ligand orbitals lay higher in energy than the
correlated ones $(\Delta > 0 )$ and the occupation is one electron
per site.

For zero hybridization two regimes can be distinguished. Depending on
the on--site energy of the ligand orbitals the ligand level enters
the lower Hubbard band or not. In the first case a
metal--to--insulator and in the second case an insulator--to--metal
transition takes place with increasing hybridization. Both forms may
exist in a range of the Coulomb interaction $U$.  The former occurs
for $U > 2 \, U_c$ and the latter for $U < 2\, U_c$.

As known from the exact solution of the Hubbard model the critical 
Coulomb repulsion $U_c$ vanishes in one dimension. We confirmed the 
absence of a self--doping transition in that case by an exact 
diagonalization study. In higher dimensions lattice symmetries and 
hopping matrices which lead to a nested Fermi surface also give 
$U_c = 0$ in the corresponding Hubbard model. In these cases the
self--doping transition again should not exist. We confirmed the 
prediction by means of an unrestricted Hartree--Fock approximation. 
Moreover, we shortly discussed the opening of a hybridization gap in 
the regime $U > 2 U_c$ and $\Delta \ll W$.
 
In the case of $U_c \neq 0$\/ the resulting picture is less conclusive. 
Both, the slave--boson and the alloy--analog approach confirm the 
existence of a self--doping transition. A restriction to a limited $U$\/ 
regime is not found here. Moreover, the transition from a metal to an 
insulator for values $U > 2 \, U_c$\/ and $\Delta \ll W$ does not occur.

One should appreciate that one is dealing here with a rather
intricate problem. The number of different approximation schemes for
investigating the Mott-Hubbard transition for finite Coulomb
repulsion is rather limited. To explore the possibility of a
self--doping transition one must be able to deal with the difficult
parameter regime $U \approx U_c$. Moreover, in addition to the
many--body Coulomb term also the hybridization term has to be
treated. The approximations applied here do not contradict the
existence of an insulator--to--metal transition with increasing $V$
in a restrictive parameter space for $U$. At present it is not clear
which of the findings are due to a particular approximation made,
i.\/e., slave--boson mean--field or alloy analogy and which ones are
independent of it. To resolve this problem further investigations
are required.

\end{multicols}

\widetext

\section*{Acknowledgement}

We thank C. Lehner for continuing discussions. This work has been
done during a visit of two of the authors (HAT and TY) to the MPI
PKS, Dresden, whose hospitality and support are gratefully
acknowledged.

\end{document}